\documentclass[%
 reprint,
 amsmath,amssymb,
 aps,nofootinbib,
article,]{revtex4-2}
\usepackage{subcaption}
\newcommand{\be}{\begin{equation}}
\newcommand{\ee}{\end{equation}}
\newcommand{\bea}{\begin{eqnarray}}
\newcommand{\eea}{\end{eqnarray}}

\usepackage[pdftex, pdftitle={Article}, pdfauthor={Author}]{hyperref}
\usepackage{sidecap}

\newcommand\aap{Astron. Astrophys.}

\newcommand\apjs{Astrophys. J. Suppl.}
\newcommand\jcap{J. Cosmol. Astropart. Phys.}
\newcommand\mnras{Mon. Not. Roy. Astron. Soc.}

\newcommand\physrep{Phys. Rep.}

\newcommand\pasj{Publ. Astron. Soc. Jpn.}

\usepackage{xcolor}
\usepackage{graphicx}
\usepackage{dcolumn}
\usepackage{caption}
\usepackage[utf8]{inputenc}
\usepackage{multirow}
\usepackage{mwe}
\usepackage{bm}
\usepackage [utf 8]{inputenc }
\usepackage {float}

\usepackage{ulem}
\newcommand\lftsout{\bgroup\markoverwith{\textcolor{olive}{\rule[0.5ex]{2pt}{2pt}}}\ULon}

\usepackage[bottom]{footmisc}

\begin{document}

\title{Impact of baryonic feedback on HSC Y1 weak lensing non-Gaussian statistics}

\author{Daniela Grandón$^{1,2}$}
\email{d.i.grandon.silva@math.leidenuniv.nl}

\author{Gabriela A. Marques$^{3,4}$}

\author{Leander Thiele$^{5,6}$}

\author{Sihao Cheng$^{7}$}

\author{Masato Shirasaki$^{8,9}$}
\author{Jia Liu$^{6}$}

\affiliation{$^{1}$Mathematical Institute, Leiden University, Snellius Gebouw, Niels Bohrweg 1, NL-2333 CA
Leiden, The Netherlands}
\affiliation{$^{2}$Grupo de Cosmología y Astrofísica Teórica, Departamento de Física,
FCFM, Universidad de Chile, Blanco Encalada 2008, Santiago, Chile}
\affiliation{$^{3}$Fermi National Accelerator Laboratory, Batavia, IL 60510, USA}
\affiliation{$^{4}$Kavli Institute for Cosmological Physics, University of Chicago, Chicago, IL 60637, USA}
\affiliation{$^{5}$Department of Physics, Princeton University, Princeton, NJ 08544, USA}
\affiliation{$^{6}$Center for Data-Driven Discovery, Kavli IPMU (WPI), UTIAS, The University of Tokyo, Kashiwa, Chiba 277-8583, Japan}
\affiliation{$^{7}$Institute for Advanced Study, 1 Einstein Dr, Princeton, NJ 08540, USA}

\affiliation{$^{8}$National Astronomical Observatory of Japan (NAOJ),
National Institutes of Natural Sciences, Osawa, Mitaka, Tokyo 181-8588, Japan}

\affiliation{$^{9}$The Institute of Statistical Mathematics, Tachikawa, Tokyo 190-8562, Japan}

\date{\today}

\begin{abstract}
  Baryonic feedback is a major systematic in weak lensing cosmology. Its most studied effect is the suppression of the lensing power spectrum, a second-order statistic, on small scales. Motivated by the growing interest in statistics beyond the second order, we investigate the effect of baryons on lensing non-Gaussian statistics and the resulting biases in the matter clustering amplitude $S_8 = \sigma_8\sqrt{\Omega_m/0.3}$. We focus on the Subaru Hyper Suprime-Cam Year 1 (HSC-Y1) data which, with its high source number density, closely resembles those expected from the upcoming Euclid and Rubin LSST. We study four non-Gaussian statistics --- peak counts, minimum counts, the probability distribution function, and the scattering transform --- in addition to the usual power spectrum. We first estimate the biases in $S_8$ using mock observations built from the IllustrisTNG and BAHAMAS hydrodynamical simulations and theoretical models built from dark matter-only simulations. We find up to $1\sigma$ bias in $S_8$ when the smallest scales (2 arcmin) and the highest feedback level are considered. 
  We then analyze the HSC-Y1 data and compare the $S_8$ obtained for each statistic with different smoothing scales or scale cuts. As we expect that baryons mostly affect the small scales, comparing the results obtained from including and excluding small scales can indicate the level of impact from baryons. With HSC data, we find only minor ($\leq0.5\sigma$) differences in $S_8$ for all statistics, even when considering very small scales (2 arcmin). Our results suggest that the effect of baryons is insignificant at the level of HSC-Y1 down to 2~arcmin for all statistics examined here, or it is canceled by other scale-dependent systematics.

\end{abstract}

\maketitle

\section{\label{sec:intro} Introduction}


Weak gravitational lensing (WL) is a powerful tool for scrutinizing the standard cosmological model, known as the $\Lambda$ Cold Dark Matter ($\Lambda$CDM) model. This phenomenon generates small distortions of the observed shapes of background galaxies caused by the gravitational influence of foreground large-scale structure (LSS) \cite{BartelSchneid}. In addition to serving as a sensitive indicator of the total density of matter in the Universe, when analyzed tomographically, WL offers a unique capability to trace the evolution of structural growth.

In recent years, Stage-III cosmic shear surveys\footnote{Definition introduced by the Dark Energy Task Force report \citep{albrecht2006report}.} such as the Dark Energy Survey (DES)~\cite{DES:2016jjg, DES:2017, DES:2021bvc}, Kilo Degree Survey (KiDS) \cite{KiDS:2015vca, KiDS:2020suj}, and the Subaru Hyper Suprime-Cam (HSC) \cite{HSC2017} have achieved great precision in constraining cosmological parameters. The conventional approach in weak lensing analyses typically involves using the power spectrum or two-point correlation function of galaxy shapes. While these summary statistics can fully describe Gaussian random fields, such as the cosmic microwave background (CMB), they are unable to capture the non-Gaussian features present in the late-time random cosmic fields. Because of physical processes such as the gravitational collapse of structures, non-Gaussianities also emerge in the lensing field, and hence statistical tools that capture this additional information are needed. For this reason, non-Gaussian statistics, also known as higher-order statistics (HOS), have gained significant attention in constraining cosmological parameters for weak lensing \cite{Zurcher:2020dvu, 2023MNRAS.525..761Z, DES:2021epj, Gabriela2023, Thiele:2023gqr}. Through Fisher forecast analysis, it has been indicated that non-Gaussian statistics can provide improved statistical power in cosmological constraints compared to relying solely on the weak lensing power spectrum \cite{2021MNRAS.505.2886B, Coulton:2018ebd, Coulton:2019enn, Liu:2018pdf, Liu:2013yna, Thiele:2020pdf, Marques:2018ctl, 2023A&A...675A.120E,Cheng_2021}. Furthermore, the application of non-Gaussian statistics to real data has demonstrated their effectiveness in constraining cosmological parameters (e.g. \cite{Liu:2014fzc,marques2023cosmological, Thiele:2023gqr}). 
As these tools capture information beyond the linear regime, they have become useful tools for studying the impact of astrophysical systematics and constraining the mass of the neutrinos \cite{Marques:2018ctl, Ajani:2020dvu, Coulton:2018ebd, 2023A&A...671A..17A}. As an example, in \cite{harnois2022cosmic,Martinet:2020omp,Coulton:2019enn,2013MNRAS.434..148S}, it has been shown that the non-Gaussian statistics can be impacted by systematic effects in different ways, and thus a detailed analysis of their sensitivity to these systematics could provide crucial guidance for developing future pipelines in weak lensing. 



One astrophysical systematic that impacts the WL signal is the baryonic feedback. This corresponds to astrophysical processes that modify the distribution of matter inside haloes, reshaping the gravitational potentials where WL occurs \cite{Chisari:2019tus, Chisari:2018prw, Sunseri:2022txp}.  As a result, it has been shown that baryons suppress matter clustering on intermediate to small scales, mainly driven by the feedback of AGN \cite{2011MNRAS.415.3649V}. The opposite affect can be observed in sufficiently small scales of $k\sim10 \, \text{h/Mpc}$, where star formation and gas cooling dominate, producing a stronger matter clustering. Unfortunately, these feedback mechanisms encompass an extensive list of astrophysical phenomena that currently lack comprehensive theoretical models, which makes their analytical treatment challenging. 

Given the expectation that baryonic physics can potentially bias cosmological constraints, various approaches have been adopted by the community to mitigate such biases. 
Stage-III surveys have applied severe scale cuts to mitigate potential biases due to the unmodelled baryonic feedback at small scales \cite{DES:2017myr, DES:2021wwk, Dalal:2023olq}, but such strategies reduce statistical power and would need to become even more restrictive for Stage-IV.
To reduce the need for scale cuts, various methods for incorporating baryonic uncertainty into the theoretical model have been employed.
These include parametrization of baryonic effects on the matter power spectrum \cite{Mohammed:2014lja, Arico2021}, or principal component analysis parametrization of the effects of baryons based on hydrodynamical simulations, whose parameters can be marginalized over in order to mitigate biases \cite{Eifler:2014iva}. Another approach consists of displacing dark matter particles in N-body simulations that can mimic the effect of baryons inside haloes \cite{Schneider:2015wta, Schneider2019, Lee:2022jyg, Weiss:2019jfx, Arico2021a}, also known as baryonification.

The efficacy of the described modeling approaches has typically only been rigorously evaluated for matter power spectra.
How well they work for the non-Gaussian statistics is an open question.
An important step towards resolving this issue is to develop intuition for the severity of baryonic biases in non-Gaussian statistics.
Developing such intuition is the primary purpose of this work.
In order to keep the presentation concise, we restrict ourselves to some of the more popular non-Gaussian statistics,
namely peak and minimum counts, the probability distribution function (PDF), and the scattering transform (ST).
These summaries have interesting differences in which features of the convergence map they retrieve information from
and should therefore provide a representative sample to elucidate the baryonic effects.
Given the heightened interest in the matter clustering amplitude, $S_8 \equiv \sigma_8\sqrt{\Omega_m/0.3}$,
we focus on this parameter.
It should be noted that some of the most interesting applications of non-Gaussian statistics may be for
parameters and model extensions beyond $S_8$ (which is optimized for two-point inference).
Since theoretical priors for baryonic feedback scenarios are poorly defined, we begin tackling the problem by making reference to
a representative subset of hydrodynamical simulations.
Using data vectors corrected for baryonic effects according to the simulations, we identify how information on
different angular scales translates into biases on $S_8$ when inference is performed using the various non-Gaussian statistics.
We then extend the analysis to real data from HSC-Y1, concentrating on \emph{relative} shifts in $S_8$ posteriors.
As HSC is the deepest Stage-III galaxy weak lensing survey conducted to date,
analyzing non-Gaussian statistics of HSC data provides a great avenue to delve into the impact of baryonic effects.
This serves as a crucial stepping stone towards the upcoming data from the Legacy Survey of Space and Time (LSST) of the Vera C. Rubin Observatory,
enabling us to enhance our understanding of systematic challenges of future weak lensing surveys.

This paper is structured as follows. In Section \ref{sec:simsdata} we present the N-body simulations from which we build our weak lensing mocks and likelihood pipeline, along with the cosmological hydrodynamical simulations that account for the baryonic feedback. Later, we describe the setup of HSC year 1 data. In this section, we assess our baryonic feedback strategy. In Section \ref{sec:method} we describe the summary statistics we employ in this work. Section \ref{sec:inference} presents the likelihood shape, and the parameter inference specifications. In Section \ref{sec:results} we show our results on the impact of baryons on the summary statistics and cosmological constraints of $S_8$, to finally compare our results with real data cosmological analysis. We summarize our results and main conclusions in Section \ref{sec: conclusions}.

\section{\label{sec:simsdata1} HSC Y1 data and simulations}
In our analysis, we utilize a set of weak lensing convergence (mass) maps that have been reconstructed from cosmological N-body and hydrodynamical simulations. The simulations employed in this paper  incorporate specific features of the HSC Y1 dataset. Therefore, we first provide a description of the HSC Y1 data, followed by a presentation of the N-body simulations and hydrodynamical simulations. Finally, we conclude this section with a step-by-step guide for our mock production and baryonic feedback strategy.

\subsection{\label{sec:simsdata}HSC-Y1 real data}
To investigate the impacts of different analysis choices on the shifts in the inferred $S_8$, we use the HSC first data release (Y1) shear catalog \citep{2018PASJ...70S..25M}. This catalog is based on observations conducted between March 2014 and April 2016 using the Subaru Hyper Suprime-Cam in five broad-bands, $grizy$. Conservative cuts are applied to select galaxies with reliable shape measurements ($S/N \geq 10$ and $i < 24.5$), resulting in a sample that spans 136.9~$\deg^{2}$ across 6 individual patches of the sky.\

The redshifts of the source galaxies are determined from HSC five broadband photometry using multiple independent codes~\citep{tanaka2018photometric}. For our analysis, we employ three tomographic bins, $0.3<z_{\rm best}<0.6$, $0.6<z_{\rm best}<0.9$ and $0.9<z_{\rm best}<1.2$, where $z_{\rm best}$ denotes the best-fit photo-$z$ determined by the \texttt{MLZ} code. This redshift range is selected to align with the accuracy range defined by the HSC team, and within which we observe no indications of unmodeled effects in our mocks, redshift miscalibration, or other systematic issues in the real data (see further discussion in \cite{Thiele:2023gqr, Gabriela2023, Dalal:2023olq}).\
 
We generate pixelized shear maps using a regular flat grid with a pixel size of 0.88 arcmin, taking into account the correction for shear responsivity and biases, as outlined in \cite{2018PASJ...70S..25M}. The shear maps are smoothed using the Gaussian kernel $W_{\rm G}$ defined as 
    \begin{equation}\label{C5_GaussianKernel}
      W_{\rm G}(\theta) = \frac{1}{2\pi\theta_{s}^2}\exp{\bigg(-\frac{\theta^2}{2\theta_{s}^2}}\bigg),
    \end{equation}
where we consider the smoothing scales of $\theta_s= \{2, 5, 8\}$ arcmin. Finally, we reconstruct the convergence maps following the Kaiser--Squires inversion method \citep{kaiser1993mapping}. For additional details on the map-making procedure, we refer the reader to \cite{Gabriela2023}.

\subsection{HSC-Y1 simulations}

The parameter inference analysis is based on two sets of N-body simulations, customized to incorporate specific features of the HSC Y1 dataset. These adaptations account for various factors, including shape noise levels, variations in the lensing weight, uncertainties in image calibration, the spatial inhomogeneity of source galaxies, uncertainties in redshift distribution, and considerations for survey geometry.

To construct an emulator that generates predictions for the summary statistics, we employ the cosmology-varied simulations introduced in \cite{Shirasaki:2019wxk}. This suite encompass 100 cosmological models within the $\Omega_m - \sigma_8$ parameter space, spanning the ranges $\Omega_m \in [0.1, 0.7]$ and $S_8 \in [0.23, 1.1]$. For each cosmological model, there are 50 ray-tracing realizations of the underlying density field.\

Furthermore, we utilize 2268 mock realizations of the HSC-Y1 shape catalogs to estimate our covariance matrix. This suite of simulations is constructed based on 108 quasi-independent full-sky $N$-body lensing simulations presented in \cite{Takahashi:2017hjr}, with the flat-$\Lambda$CDM model cosmology consistent with the best-fit result of the Wilkinson Microwave Anisotropy Probe (WMAP) nine-year data \citep{WMAP9}: $\Omega_{\text{b}}$ = 0.046, $\Omega_{\text{m}}$ = 0.279, $\Omega_\Lambda$ = 0.721, $h$ = 0.7, $\sigma_8$ = 0.82, and $n_s$ = 0.97. Finally, we adapted these simulations to replicate the properties of the real data, following the steps outlined in Section 3.1 of \cite{Gabriela2023}. We reconstruct the convergence maps using the same procedure as for real data, including the chosen redshift bins and smoothing scales.

\subsection{Hydrodynamical simulations}

In order to model baryonic feedback processes, a state-of-the-art strategy involves the use of hydrodynamical simulations, wherein baryonic effects are mimicked by sub-grid models. In this work, we utilize BAryons and haloes of MAssive Systems (BAHAMAS, \cite{McCarthy:2016mry,Mccarthy:2017yqf}) and  IllustrisTNG (TNG300-1) \cite{Vogelsberger:2014dza,Osato:2020sxo} simulations to reconstruct weak lensing convergence maps. Each of these simulations considers specific calibration strategies, reproducing a subset of observables (e.g., star formation history, stellar-to-halo-mass relation) to a reasonable degree within the measurement uncertainties. Therefore, by using the two hydrodynamical simulations, we can explore and compare the extent to which different implementations of baryonic processes can impact our cosmological constraints. 
As described below, the simulations are used to compute ratios of full-physics vs.\ gravity-only summary statistics.
Hence, the requirement on the mocks are somewhat relaxed since slight model-misspecification compared to the HSC-Y1 mocks
cancels to leading order when computing the ratio.
\subsubsection{BAHAMAS}

BAHAMAS is a set of cosmological hydrodynamical simulations of box size 400 $h{^{-1}}$ Mpc. They are designed to replicate the observed stellar and hot gas properties of massive haloes \cite{Mccarthy:2017yqf}. As the AGN feedback is considered one of the most influential baryonic feedback mechanisms in LSS \cite{2011MNRAS.415.3649V}, we utilize three BAHAMAS runs in which AGN heating temperature is varied. The purpose of varying this temperature is to encompass the scatter observed in the gas fraction for galaxy groups in X-ray observations. For the fiducial model, the subgrid parameters are calibrated such that the efficiencies of stellar and AGN feedback match the observed amplitude of hot gas fraction–halo mass relation of groups and clusters, and galaxy stellar mass function (for $M_{*}> 10^{10} \text{M}\textsubscript{\(\odot\)}$). For the other two models, the AGN heating temperature is raised and lowered by 0.2 dex. We refer to those models as high and low AGN, respectively.  We utilize BAHAMAS simulations at the WMAP 9-yr cosmology $\Omega_{\text{b}}$ = 0.0463, $\Omega_{\text{m}}$ = 0.2793, $h$ = 0.7, $\sigma_8$ = 0.821, and $n_s$ = 0.972 \cite{WMAP9}, from which we construct HSC-Y1-like weak lensing convergence maps for 60 source redshifts up to $z_s=3$. Each map covers an area of $5\times5$ deg$^2$ of the sky and contains $340^2$ pixels.

\subsubsection{IllustrisTNG}
The TNG300-1 simulations are a set of cosmological and large-scale hydrodynamical simulations of box-size 205 $h^{-1}$ Mpc \cite{Vogelsberger:2014dza, 2019ComAC...6....2N} at the \textit{Planck} 2016 cosmology: $\Omega_{\text{b}}$ = 0.0486, $\Omega_{\text{m}}$ = 0.3089, $h$ = 0.6774, $\sigma_8$ = 0.8159, and $n_s$ = 0.9667 \cite{Planck2015}. Sub-grid prescriptions are aimed at modeling black hole feedback, thermal and kinetic AGN feedback, stellar evolution, chemical evolution, galactic winds, and magnetic fields \cite{Pillepich:2017fcc}, among others. In this work, we use the already existing $\kappa$TNG convergence maps presented in \cite{Osato:2020sxo}. These $1024^2$ pixels maps cover $5\times5$ deg$^2$ of the sky for 40 source redshifts up to $z_s=2.6$, obtained from random rotations, translation, and flips of the snapshots.\\

\subsubsection{Treatment of the Hydrodynamic mocks}
We build hydrodynamic mocks based on the BAHAMAS and $\kappa$TNG light cones. We refer to the baryonic feedback models as: BAHAMAS low-AGN, BAHAMAS fiducial-AGN, BAHAMAS high-AGN, and IllustrisTNG. For each simulation, we have 10,000 realizations of convergence maps at source redshifts. All of them pose a dark-matter-only counterpart, based on the same corresponding simulation and initial conditions.\ 

Our methodology proceeds as follows:

\begin{enumerate}
\item Initially, we assign weights to the lightcone simulations based on the HSC Y1 galaxy source redshift distribution and sum them along the line-of-sight. We adopt the tomographic redshift bins outlined in Section \ref{sec:simsdata}, utilizing three bins that cover the same range as the real data, i.e., $0.3 < z < 1.2$. 
    \item In order to mimick the HSC-Y1 shape noise, we add noise to each pixel, whose value is drawn from a Gaussian distribution centered at zero, with variance 
    \begin{equation}\label{C5_shapenoise}
        \sigma^2 = \frac{\sigma_{e}^2}{ n_{g}^{\rm eff} A_{\rm pix}},
    \end{equation}
    where $\sigma_{e}\sim 0.28$ is mean intrinsic ellipticity of galaxies, $n_{g}^{\rm eff}$ is the effective galaxy number density 5.14, 5.23, 3.99 arcmin$^{-2}$ for each tomographic bin, and $A_{\rm pix}$ is the solid angle of a pixel in units of arcmin$^{-2}$.\
    
    \item Smoothing convergence maps is a standard procedure to reduce the shape noise per pixel, and to exploit the features encoded in the maps \cite{Weiss:2019jfx}. In our case, this method is useful to investigate the scale at which the baryonic effects can be sufficiently mitigated. Therefore, we apply the Gaussian smoothing filter in Eq.~\ref{C5_GaussianKernel} to the mocks. We consider the smoothing scales $\theta_s= \{2, 5, 8\}$ arcmin. 
    
    \item \textbf{Data vector and baryon injection:}
The method we employ to account for the presence of baryons proceeds as follows: 

\begin{enumerate}
\item We measure the summary statistics on convergence maps that include baryonic feedback, which we denote as $\text{N}^{\text{H}}$ for the peak and minimum counts, $\text{PDF}^{\text{H}}$ for the probability distribution function,  \{$\text{s}^{\text{H}}_1$ , $\text{s}^{\text{H}}_2$\} for the scattering transform coefficients, and $C^{H}_{\ell}$ for power spectrum. We repeat on the corresponding dark matter-only counterparts, where the results are denoted with the upper label 'DM'.
\item We divide the two data vectors, which generates the ratios $\text{N}^{\text{H}}/\text{N}^{\text{DM}}$, $\text{PDF}^{\text{H}}/\text{PDF}^{\text{DM}}$, $\text{s}^{\text{H}}_1/\text{s}^{\text{DM}}_1$, $\text{s}^{\text{H}}_2/\text{s}^{\text{DM}}_2$ and $C^{\text{H}}_{\ell}$/$C^{\text{DM}}_{\ell}$. These results quantify the impact of baryons on weak lensing non-Gaussian statistics and power spectrum.

\end{enumerate}

In order to propagate the effect of baryons into cosmological constraints, we inject baryons to the DM-only data vectors as follows

\begin{equation}
    \text{DV}^{\text{H}} = \text{DV}^{\text{DM}} \frac{\langle N^{\text{H}} \rangle}{\langle N^{\text{DM}} \rangle},    
\end{equation}
where we consider the ratio as a correction factor (with the peak/minimum counts ratio as an example) and the angular brackets denote average over 10.000 realizations. This step-by-step methodology is repeated for the different analysis choices: summary statistics, tomographic bins and smoothing scales.

\end{enumerate}

\section{\label{sec:method} Non-Gaussian statistics and power spectrum}

In this work, we focus on the power spectrum, peak counts, minimum counts, PDF and scattering transform coefficients, to study the convergence field.

\subsection{\label{sec:peakmin} Peaks and Minimum counts}
Peak statistic has been studied in the literature extensively and hence it is positioned as one of the most popular non-Gaussian statistics to study the WL convergence field \cite{Liu:2013yna, Coulton:2018ebd}. We employ the definition of peaks which is the counting of pixels in the smoothed map whose (convergence $\kappa$) value is higher than their surrounding eight pixels. It has been shown that high peaks trace the massive haloes in the line of sight, whereas low peaks trace the superposition of smaller haloes \cite{Liu:2016xjb}. Conversely, minimum counts correspond to pixels with lower values compared to their eight neighbors, tracing underdense regions and probing information complementary to the peak counts alone \cite{Coulton:2019enn}. We measure the peaks in the
linear signal-to-noise ratio $\kappa/\langle\sigma(\kappa) \rangle$ with 19 equally spaced bins from -4 to 4. It is important to note that the average $\langle\sigma(\kappa) \rangle$ refers to the average over the individual $\sigma(\kappa)$ obtained for each map realization. We remove the extreme bins for both statistics and apply further scale cuts, which
varies according to the emulator performance. For the peak counts, we utilize 14 bins in the range $-1.8 < \kappa/\langle\sigma(\kappa) \rangle < 4$, whereas for the minimum counts the same number of bins is obtained encompassing the values $-4 < \kappa/\langle\sigma(\kappa) \rangle < 1.8$.

\subsection{\label{sec:pdf} Probability distribution function (PDF)}
The probability distribution function is a summary statistic that captures the amplitude of the weak lensing signal. It is sensitive to the NG information contained in the fields and has been shown to contribute to tightening cosmological constraints \cite{Liu:2018pdf}. Analytic modeling approaches, such as the halo model \cite{Thiele:2020pdf} or large deviation statistics \cite{2021MNRAS.505.2886B, Bernardeau:2000et, Barthelemy:2020yva, 2023arXiv230709468B}, can be employed to study and interpret the PDF. In this work, the convergence PDF is built upon histogramming pixels of convergence maps, following the same method described for the peak counts and minimum counts, to enable clear comparisons in our results. It is important to note that this method differs from the PDF analysis presented in \cite{Thiele:2023gqr}, where the authors divide each map by its own standard deviation (instead of being divided by an average standard deviation, as we do in this work) and hence the inference relies on the non-Gaussian character of the map. As for peaks and minima, the PDF is calculated in 19 equally spaced signal-to-noise (S/N) bins $\kappa/\langle \sigma(\kappa) \rangle$ with 19 equally spaced bins from -4 to 4. We consider the range $-2.6 < \kappa/\langle\sigma(\kappa) \rangle < 3.6$. \subsection{\label{sec:st} Scattering Transform}
\begin{figure*}[!t]
        \centering
                \begin{subfigure}[b]{0.74\textwidth}
            \centering
            \includegraphics[width=\textwidth]{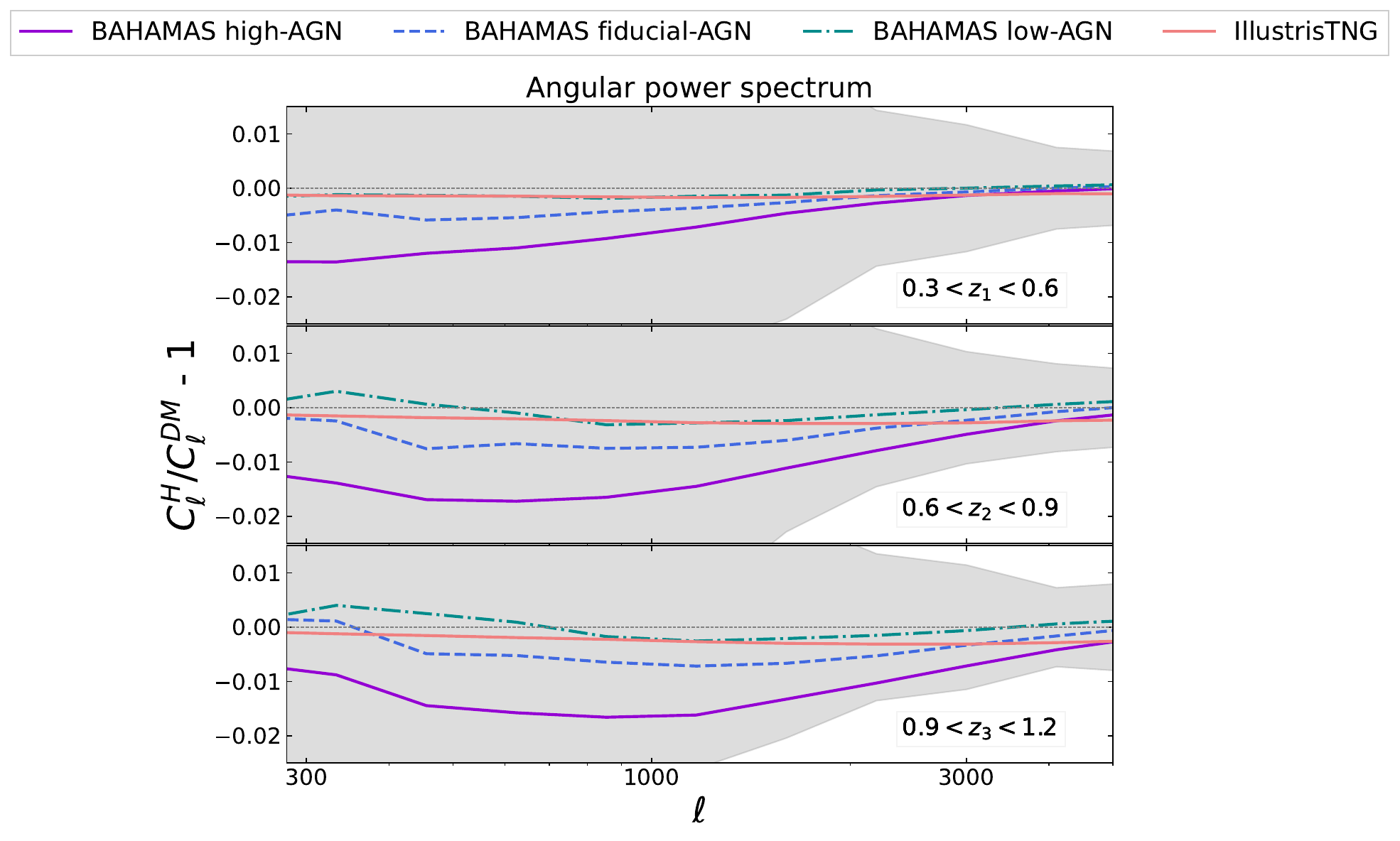}
        \end{subfigure}
        \hfill
        \vskip\baselineskip
        \begin{subfigure}[b]{0.49\textwidth}
            \centering
            \includegraphics[width=\textwidth]{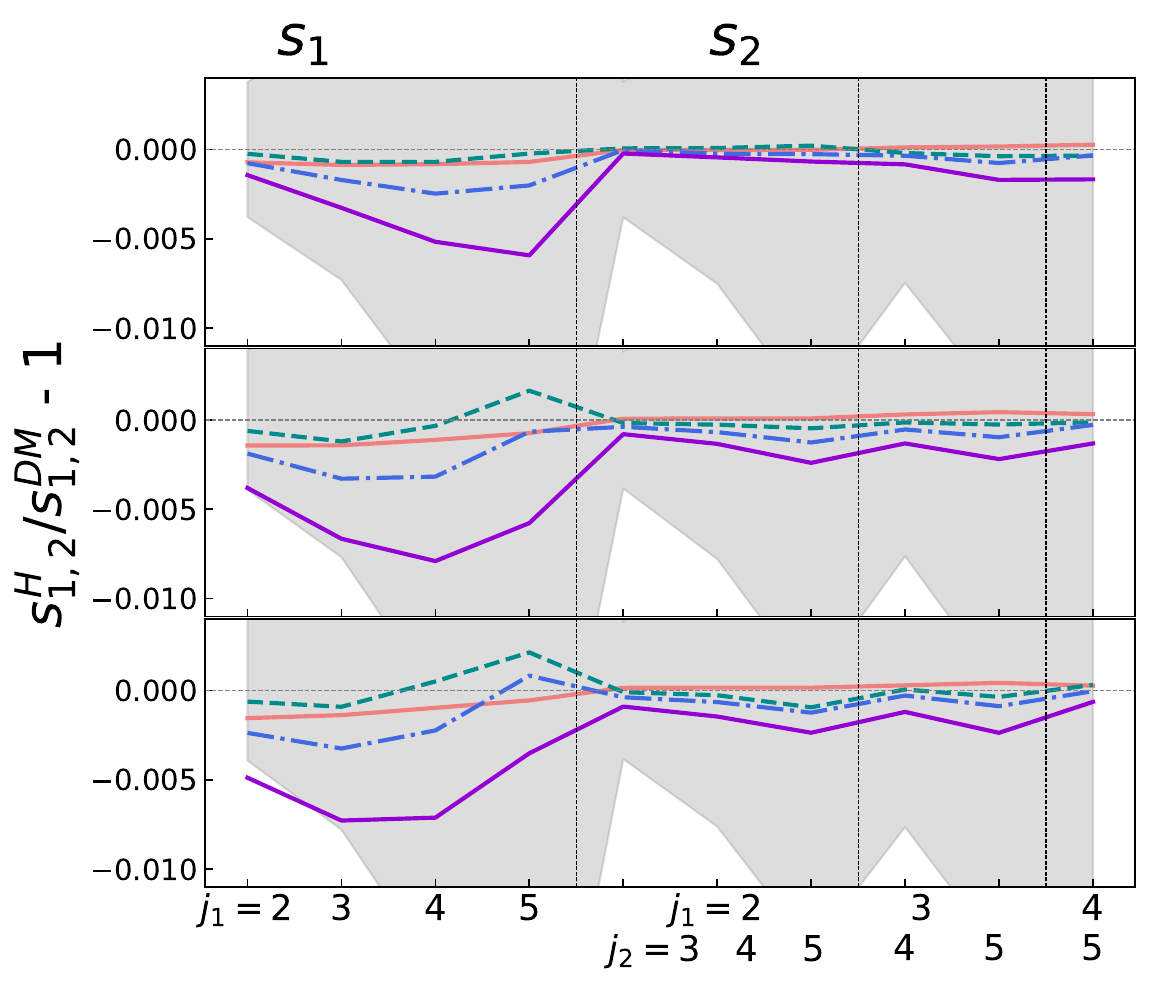}
        \end{subfigure}
        \begin{subfigure}[b]{0.475\textwidth}  
            \centering 
            \includegraphics[width=\textwidth]{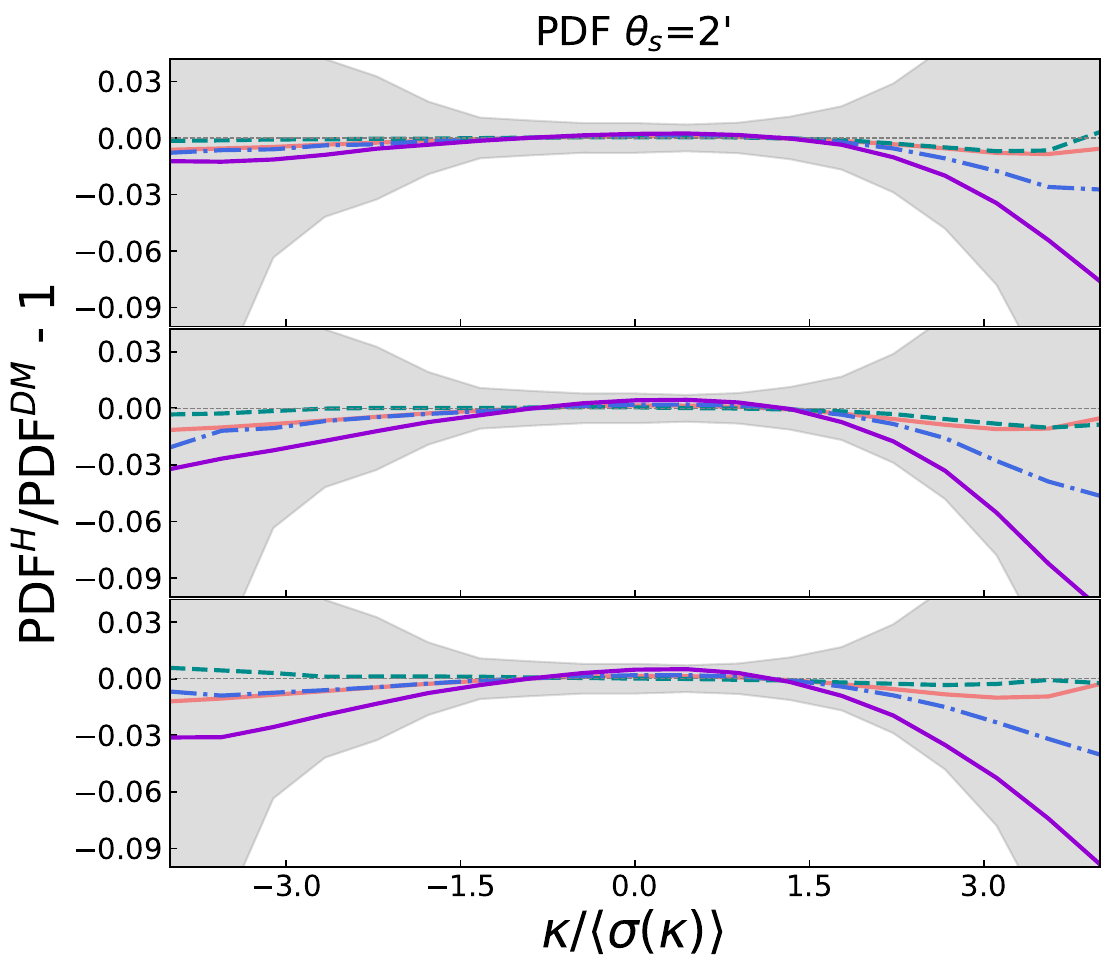}
        \end{subfigure}
  
            \caption 
        {\small The impact of baryonic feedback on the convergence power spectrum (top), scattering transform coefficients $s_1$ and $s_2$ (left) and probability distribution function (right) for $\kappa$-maps smoothed with a Gaussian kernel of $\theta_s = 2$ arcmin. We show the results for the hydrodynamical simulations BAHAMAS high-AGN (solid purple line), fiducial-AGN (dashed-dot blue line), low-AGN (dashed teal line) and $\kappa$TNG (solid pink line). The vertical axis corresponds to the fractional difference where $C_{\ell}^{\text{H}}$ corresponds to the baryon-injected data vector (same for $s^{\text{H}}_{1,2}$ and PDF$^{\text{H}}$), and $C_{\ell}^{\text{DM}}$ the dark-matter-only data vector (and also $s^{\text{DM}}_{1,2}$ and PDF$^{\text{DM}}$, based on the dark matter-only runs of each simulation). The grey-shaded region corresponds to $1\sigma$ HSC-Y1 uncertainty.} 
        \label{fig:ratios1}
                 \end{figure*}

The scattering transform (ST) is a powerful statistical tool borrowing ideas from convolutional neural networks but does not need training \cite{Bruna_2013}. It has been recently introduced to cosmology and shown to extract a similar amount of information as convolutional neural networks in the weak lensing context \cite{Cheng_2020}. It also demonstrates stability properties, a desired advantage for practical applications \cite{Cheng_2021}. For a statistically homogeneous and isotropic field, it consists of a set of coefficients ${s_1, s_2}$. 
The coefficients $s_1(j_1)$ capture the strength of fluctuations at different scales $j_1$, therefore have a similar interpretation as the power spectrum. The coefficients $s_2(j_1,j_2)$ are sensitive to localized structures and capture the non-Gaussian information in terms of sparsity and shape at different combinations of scales ($j_1, j_2$). 
More details can be found in a pedagogical paper \cite{Cheng_2021guide}. 
Partly based on wavelet transform, scales are naturally separated in the scattering transform and are labeled by the wavelet size index $j_1$. Here we use the same wavelets and definition of the scattering transform as in \cite{Cheng_2020}. The scale increases by a factor of 2 when $j_1$ increases by 1. To cover the scales of $300<\ell<1900$, we use only wavelets with $j_1=3,4$, which have central frequencies of $\ell=1200$ and $550$, respectively. As a result, for each tomographic map, we use 3 scattering coefficients in our analysis: $s_1(3), s_1(4), s_2(3,4)$. We do show coefficients with more scales in Fig.~\ref{fig:ratios1} for visualization purpose.

\subsection{Power spectrum}
We compute the pseudo-$C_\ell$ estimator on the maps using \texttt{NaMaster} public code \cite{Alonso:2018jzx}\footnote{\url{https://namaster.readthedocs.io/}}. We obtain the pseudo-power spectrum for each tomographic redshift bin in 14 logarithmically equally spaced angular multipole bins spanning the range $81 <\ell< 6580$. To assess the influence of baryonic effects, we consider three different scale cut strategies: $300 < \ell < 900$, $300 < \ell < 1900$, and $900 < \ell < 1900$. A lower multipole range $\ell<300$ is excluded due to the presence of B modes residuals coming from unmodelled PSF systematics, as indicated in \cite{hikage2019cosmology, oguri2018two}. This was further validated in our companion paper \cite{marques2023cosmological}, and in \cite{Thiele:2023gqr}. Scales beyond $\ell>1900$ are also excluded due to other unmodeled systematics \cite{hikage2019cosmology}.

  \begin{figure*}[!t]
        \begin{subfigure}[b]{0.475\textwidth}

        \centering
            \centering 
            \includegraphics[width=\textwidth]{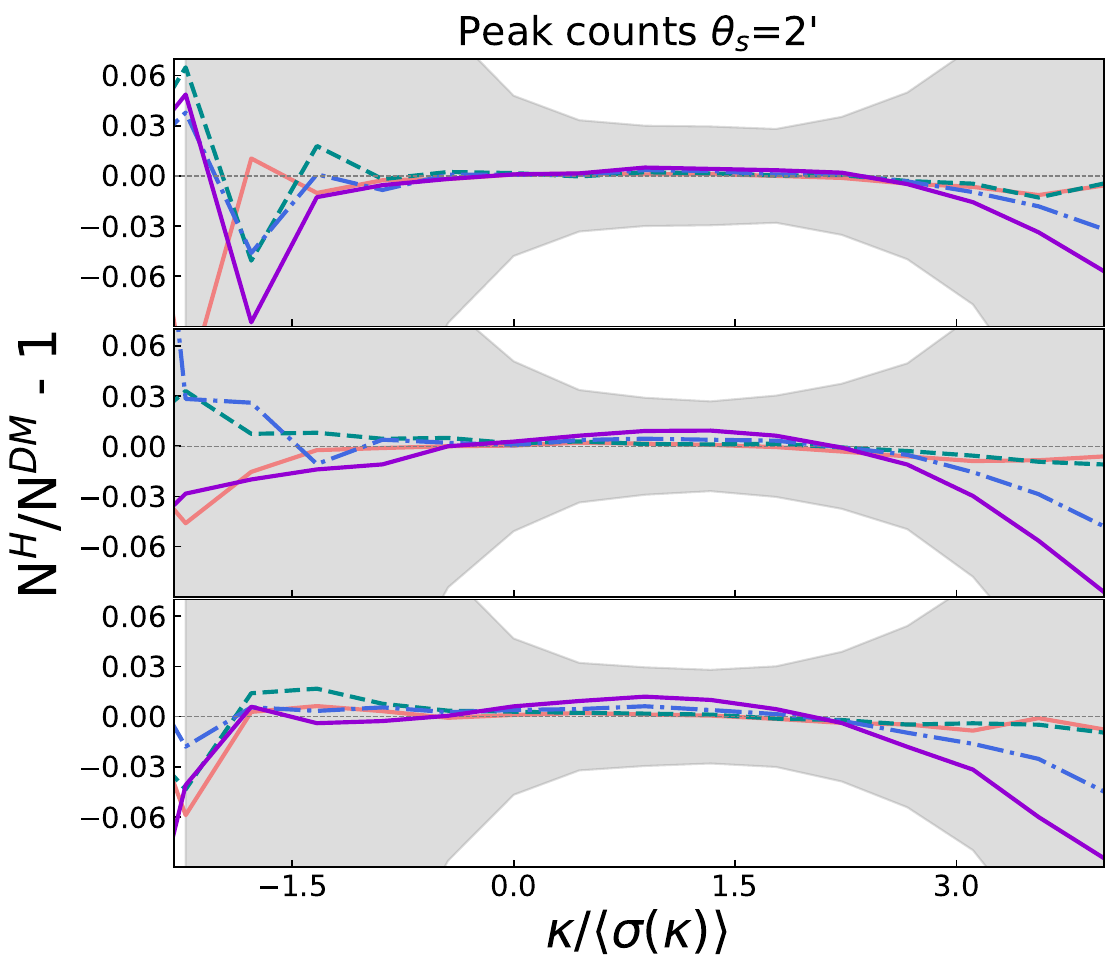}
        \end{subfigure}
        \hfill
        \begin{subfigure}[b]{0.475\textwidth}   
            \centering 
            \includegraphics[width=\textwidth]{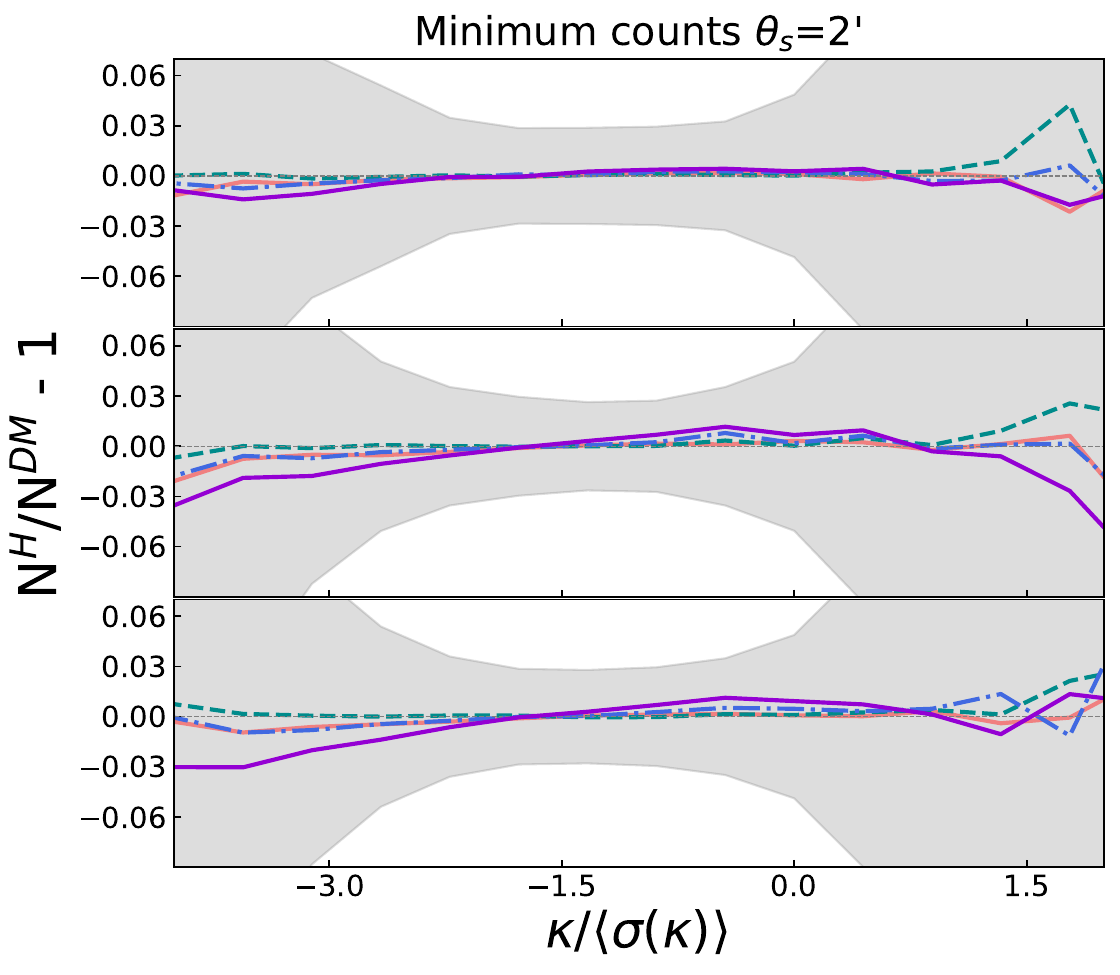}

        \end{subfigure}
        \caption 
        {\small Same as Fig.~\ref{fig:ratios1}, but for the peak counts (left), and minimum counts (right) for $\kappa$-maps smoothed with a Gaussian kernel of $\theta_s = 2$ arcmin. } 
        \label{fig:ratios2}
    \end{figure*}

\section{\label{sec:inference} Parameter inference}
To model and predict the summary statistics at arbitrary cosmologies, we train a Gaussian Processes emulator with a Radial-Basis-Function (RBF) kernel implemented in scikit-learn\footnote{\url{https://scikit-learn.org}}. The training set consists of 100 cosmological models from \cite{Shirasaki:2019wxk} and the associated summary statistics.
We emphasize that the training mocks do not contain baryonic effects.
We train multiple emulators, each of them trained to predict an individual element of the data vectors, given input cosmological parameters $S_8$ and $\Omega_m$.
In order to prevent overfitting and search for the best hyperparameters, we perform a leave-one-out cross-validation test, at which we compare the truth value with the emulator prediction. Our emulators present on average $\sim 2-3\%$ of fractional error between the prediction and truth for all statistics and smoothing scales. It is important to note that the scale cuts are decided based on preserving the emulator precision and, at the same time, preserving the maximum cosmological information from the bins. 

For the non-Gaussian statistics, we apply the data linear compression algorithm \texttt{MOPED} \cite{Heavens:1999am}, which reduces the number of bins to the number of cosmological parameters (two in this analysis, $S_8$ and $\Omega_m$) while preserving the Fisher information. With this, we can reduce the noise of the covariance matrix and Gaussianize the likelihood \cite{gatti_chang}. The compressed data vector $D^{\rm compr}$ equals
\begin{equation}
    D^{\rm compr}_\alpha = \frac{\partial D^{\mathrm T}}{\partial p_{\alpha}} C^{-1} D,
\end{equation}
where $D$ is the data vector before compression, and $\frac{\partial D}{\partial p_{\alpha}}$ is the partial derivative of the model data vector with respect to the $\alpha$-th parameter. 

Finally, we adopt a Gaussian likelihood given by

 \begin{equation}
    \mathcal{L}(\textbf{x} | \textbf{p}) \propto \exp \left(-\frac{1}{2} [\textbf{x}- \boldsymbol{\mu}(\textbf{p})]^T C^{-1} [\textbf{x}- \boldsymbol{\mu}(\textbf{p})] \right),
\label{eq:likelihood}
\end{equation}
where $\textbf{x}$ is the data vector and $\boldsymbol{\mu}$ the emulator prediction as a function of the parameters \textbf{p}. The covariance matrix $C$ is built based on the 2268 realizations of the fiducial model from full-sky N-body simulations \cite{Takahashi:2017hjr} described in \ref{sec:simsdata}. 
In order to obtain an unbiased inverse of the covariance matrix, we apply the Hartlap factor $(N_s - N_b - 1)/(N_s -1)  $\cite{Hartlap:2006kj}, where $N_s$ is the number of realisations and $N_b$ the number of bins.

We perform a Monte Carlo Markov Chain (MCMC) analysis for Bayesian inference of cosmological parameters $\Omega_m$ and $S_8$, adopting the flat priors $0.15 < \Omega_m < 0.45$ and $0.45 < S_8 < 1$. This is implemented through the public code \texttt{Cobaya} \cite{Torrado_2021, Torrado:2020dgo}.

\begin{figure*}[!t]
        \centering
                \begin{subfigure}[b]{0.77\textwidth}
            \centering
            \includegraphics[width=\textwidth]{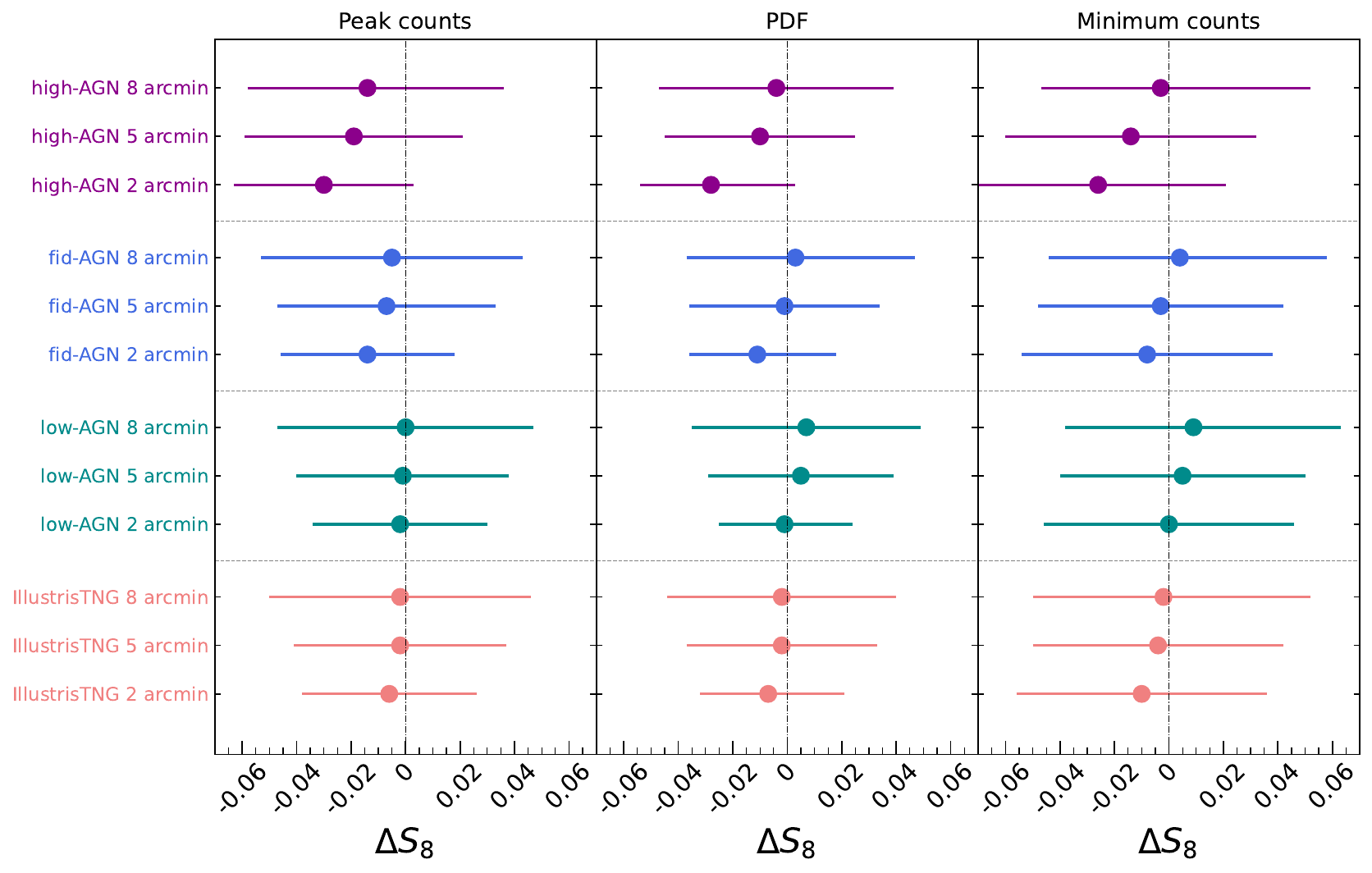}
        \end{subfigure}
        \vskip\baselineskip\hspace{15ex}
        \begin{subfigure}[b]{0.77\textwidth}
            \includegraphics[width=\textwidth]{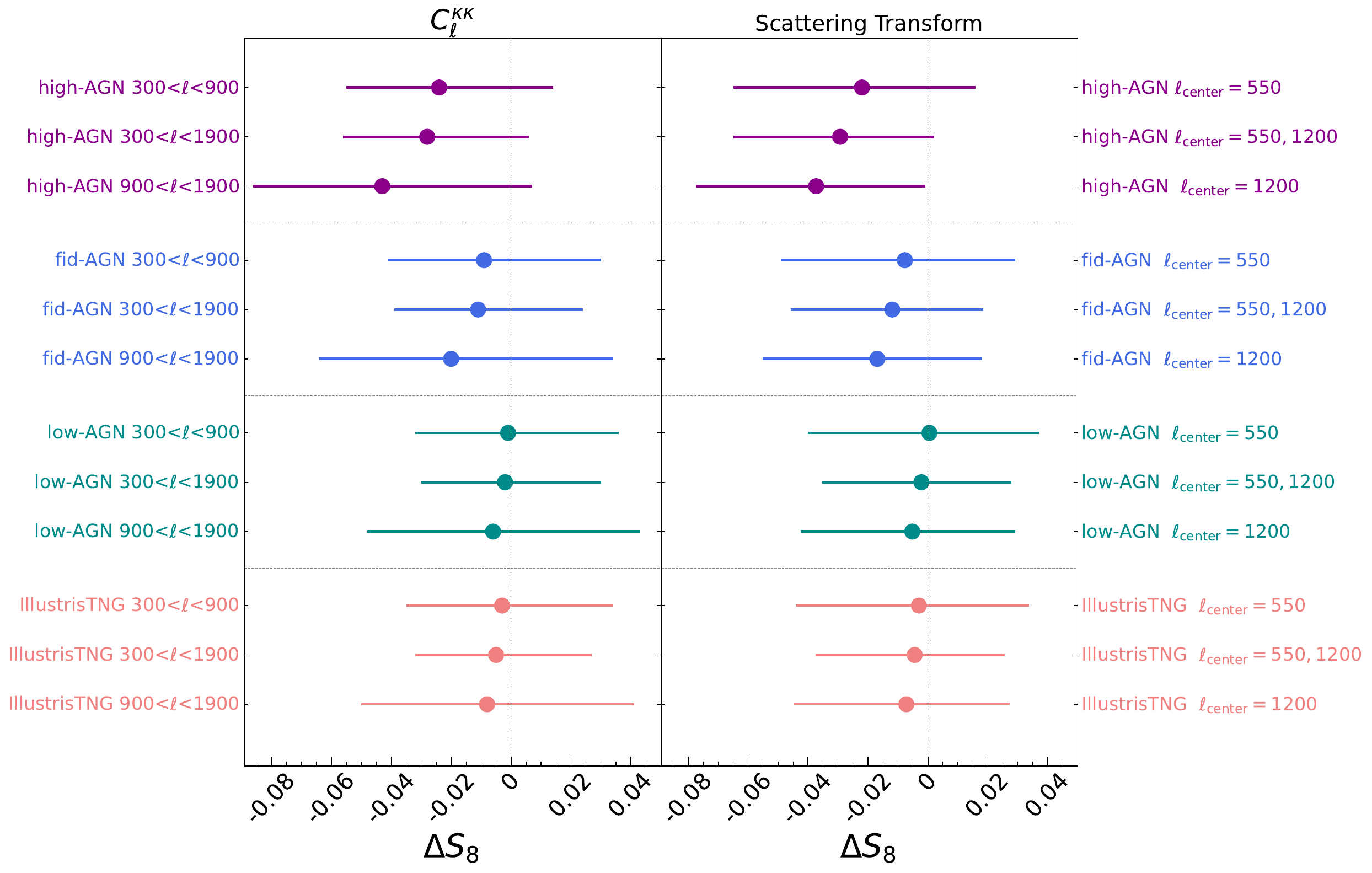}
        \end{subfigure}
  
            \caption 
        {\small The impact of baryonic feedback on $S_8$ for the non-Gaussian statistics and power spectrum pseudo-$C_{\ell}$ using mock data. We show the results for the hydrodynamical simulations BAHAMAS high-AGN (purple), fiducial-AGN (blue), low-AGN (teal) and IllustrisTNG (pink). The vertical axis corresponds to the scales considered and the horizontal axis is the $S_8$ discrepancies defined as 
     $\Delta S_8 = S_{8}^\text{H} - S_{8}^{\text{DM}}$ where `H' stands for the resulting $S_8$ from the baryon-corrected synthetic data vector and `DM' for dark matter only (no baryonic correction).} 
        \label{fig:ratios3}
                 \end{figure*}

\section{\label{sec:results} Results}

We first give an overview of the data vector-level effects of baryons, before turning to the results on inference of $S_8$ for both synthetic and observed data vectors.

\subsection{Data vectors}

The effect of baryons on the summary statistics and tomographic bins is shown in Fig.~\ref{fig:ratios1} and Fig.~\ref{fig:ratios2}. For the non-Gaussian statistics, these figures show the results for the convergence maps smoothed with a Gaussian kernel of $\theta_s=2$ arcmin. The grey shaded region corresponds to $1\sigma$ uncertainty from HSC-Y1, and the vertical axis depicts the fractional difference between baryonic physics data vectors and the dark matter only case. The main effect of baryons in the four baryonic feedback scenarios considered in this work, namely $\kappa$TNG, low-AGN, fiducial-AGN, and high-AGN  is the suppression of the structures at $\kappa/\langle\sigma(\kappa) \rangle > 1.5$ and $\kappa/\langle\sigma(\kappa) \rangle < -1.5$. This is clearly seen for the PDF, peak counts, minimum counts, and the scattering transform coefficients. The impact of these baryon scenarios exhibit variations in amplitude and scale, consistent with previous studies \cite{Chisari:2019tus}. In Fig.~\ref{fig:ratios1}, top panel, we show the fractional difference for the convergence power spectrum, where the BAHAMAS high-AGN reaches $\sim10\%$ to ~$\sim15\%$ suppression for all tomographic bins at scales $\ell \approx 300$ to $\ell \approx 1000$, with a more pronounced effect observed in the third tomographic bin $z_3$. This result is also observed for the scattering transform coefficients, $s_1(j_1)$ being sensitive to the same information contained in the power spectrum (the strength of matter fluctuations). For the power spectrum, we also observe a turn-around, meaning baryons start to enhance the the clustering of structures, which is mainly attributed to gas cooling and star formation at the smaller scales.  In all redshift bins, the effects of the BAHAMAS low-AGN and $\kappa$TNG scenarios are the weakest, being comparable between each other. 

For the smoothing scale $\theta_s=2$ arcmin, the peaks and PDF are the most impacted by baryons when considering BAHAMAS high-AGN, with suppressions in $\kappa$ of approximately ~$\sim 10\%$ in the high $\kappa$ regime; and ~$\sim 6\%$ for BAHAMAS fiducial-AGN. This means that the presence of baryons dilutes the overdense regions for the baryonic feedback scenarios considered, and thus matter is redistributed, populating underdense regions as well. This feature is captured by the supression of the negative (mainly for the PDF and minimum counts) and positive $\kappa$ tails. The three BAHAMAS scenarios exhibit a slight increase in structures in the range $0 < \kappa/\langle\sigma(\kappa) \rangle < 1.5$ for peaks and PDF, due to the material from high $\kappa$ being removed and placed in lower $\kappa$ regions under the effect of AGN feedback. 
It is important to highlight that although the deviations from zero are relatively small in comparison to the $1\sigma$ uncertainties of the HSC-Y1 data, the correlations between the bins are not depicted in the Figures \ref{fig:ratios1} and \ref{fig:ratios2}, and could potentially influence the overall parameter inference. Therefore, we proceed to investigate the total impact of such deviations in terms of the cosmological parameters.

\begin{figure*}
\centering
\includegraphics[width=0.7\textwidth]{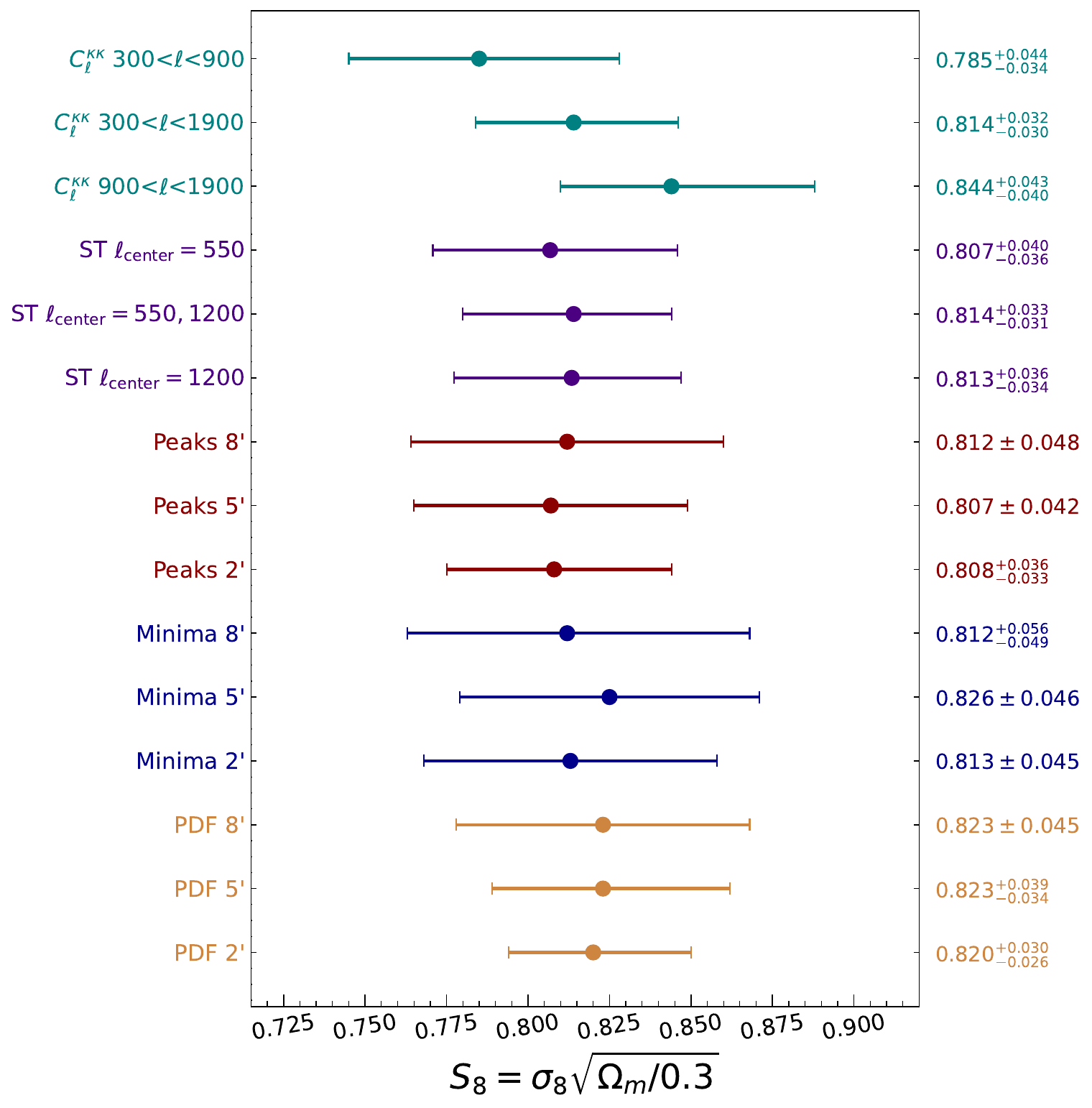}
         
  \caption{Results for HSC-Y1 real data analysis, where error bars show 68\% credible regions. The horizontal axis corresponds to $S_8 \equiv\sigma_8 \sqrt{\Omega_m/0.3}$ cosmological constraints for the power spectrum pseudo-$C_{\ell}$ (teal), scattering transform (purple), peak counts (red), minimum counts (blue), and PDF (yellow).  We report the $S_8$ median and 1$\sigma$ uncertainties at the right side of the vertical axis.}
\label{fig:realdata}   
\end{figure*} 

\subsection{Mock data: Impact on $S_8$}
In Fig.~\ref{fig:ratios3}, we show the impact of baryons on the cosmological forecasts on $S_8$. It is important to emphasize that the likelihood in Eq.~\ref{eq:likelihood} incorporates data vectors corrected by the same effects observed in Fig.~\ref{fig:ratios1} and Fig.~\ref{fig:ratios2}. The horizontal axis displays the shift $\Delta S_8 = S_{8}^{\text{H}} - S_{8}^{\text{DM}}$, where $S_8^\text{H}$ denotes posterior median inferred from the baryon-corrected data vectors, $S_8^{\text{DM}}$ denotes the results obtained from dark matter-only data vectors. 

For the power spectrum with scale range $900<\ell<1900$, we observe that the impact of baryons leads to a bias of $-0.93\sigma$ in the high-AGN baryonic feedback scenario, while the bias is $\sim -0.16\sigma$ for TNG and low-AGN scenarios. For large scales $300<\ell<900$, the strongest feedback produces $-0.7\sigma$ bias. The best constraining power for power spectrum is achieved when considering all data points, namely  $300 < \ell < 1900$. In this case, biases range from $-0.17\sigma$ until $-0.9\sigma$ depending on the baryon case. Therefore, even if we aim at extending scales up to $\ell=1900$ to achieve better precision, the power spectrum is not strongly affected by baryons and hence we can preserve accuracy even if we include baryonic feedback in the analysis.\\ 

On the other hand, the non-Gaussian statistics peak counts for $\theta_s=2$ arcmin and scattering transform with $\ell_{\text{center}}=1200$ exhibit $\sim -0.94\sigma$ biases for high-AGN, similar to power spectrum. For the same analysis choices, the PDF is the statistic most impacted, with a $-0.98\sigma$ bias; and minimum count the most robust against baryonic feedback ($-0.55\sigma$), consistent with Ref.~\cite{Coulton:2019enn}.
In general lines, our results indicate that a 5 arcmin smoothing scale is enough to preserve accurate constraints on $S_8$ for HSC-Y1, including all the different baryon models covered in this work.

\subsection{Real data results}

We repeat our inference analysis using the HSC Y1 real data vectors, rather than a synthetic data vector corrected by baryonic effects. We aim to investigate whether we observe any significant shifts in $S_8$ derived from real data when employing the same analysis choices presented for mock data, with a special interest in the impact of including the small scales. It is important to note that for this analysis baryons are no longer modeled, therefore we are not thinking of baryonic feedback as an isolated systematic. Still, we aim to further investigate the trends found with HSC Y1-like mock data, and have a wide comparison of the non-Gaussian statistics for HSC real data. 

We show the results for the HSC Y1 real data analysis in Fig.~\ref{fig:realdata}. Our results are in good agreement with previous HSC Y1 constraints, such as the results based on the power spectrum analysis $S_8 = 0.780^{+0.030}_{-0.033}$ from Ref.~\cite{hikage2019cosmology}, and the two-point correlation function constraints $S_8= 0.823^{+0.032}_{-0.028}$ from Ref.~\cite{hamana_hsc}. It is important to note that our analysis differs from these studies in many aspects, including the tomographic bins, covariance matrix prescription, the parameters considered for parameter inference, among others. Besides the aforementioned HSC Y1 two-point analyses, our results are also statistically consistent with HSC Y1 deep learning results from Ref \cite{Lu:2023dep}.

The real data results in Fig.~\ref{fig:realdata} show no statistically significant shift in $S_8$ when comparing large and small scales. For instance, the scattering transform $\ell_\text{center}=1200$ shows $0.3\sigma$ bias in $S_8$ respect to central frequency $\ell_\text{center}=550$. For the rest of the non-Gaussian statistics, this bias is also very minor. The inferred $S_8$ from power spectrum pseudo-$C_{\ell}$  with scale range $900 < \ell < 1900$ exhibits a $1\sigma$ bias towards higher $S_8$. In order to investigate this result further, we run a cosmological analysis employing a naive estimation of the $C_{\ell}$ (without considering th pseudo-$C_{\ell}$ approach). Our constraints are:  $S_{8} = 0.825^{+0.042}_{-0.046}$ for $C_{\ell}$ in the scale range $900 < \ell < 1900$, $S_{8} = 0.810^{+0.034}_{-0.030}$ for $300 < \ell < 1900$, and $S_{8} = 0.804^{+0.038}_{-0.034}$ for $300 < \ell < 900$. Therefore, the trend observed in the $S_8$ constraints from the pseudo-$C_\ell$ analysis, particularly when including small scales, becomes less significant when adopting the naive power spectrum computation. This can be attributed to the potential presence of systematics in real data, which are captured by the mixing of scales in the pseudo-$C_{\ell}$ approach. Further investigation of this phenomenon is warranted in future studies.

Finally, from Fig. \ref{fig:realdata} we observe that the constraints from the PDF for 2 arcmin smoothing scale outperform all statistics. The PDF is followed by the scattering transform with central frequency $\ell_\text{center}=550, 1200$ and the peak counts 2 arcmin, with constraining power similar to power spectrum pseudo-$C_{\ell}$ $300 < \ell < 1900$. Hence, a combined analysis of these non-Gaussian statistics can provide tighter cosmological constraints. It's noteworthy that while both \cite{Thiele:2023gqr} and \cite{Gabriela2023} employ the same set of simulations used in this work, the results presented in Fig. \ref{fig:realdata} exhibit differences in constraining power. This divergence stems from varied analysis choices, encompassing the combination of smoothing scales considered, how the convergence maps were normalized, and other specific choices.
 
Only focusing on baryonic physics as a systematic effect, these results suggest that the impact of baryonic effects on the HSC Y1 real data is potentially sub-dominant than what is expected in simulations. However, it is essential to consider the complexity and limitations of the simulations when interpreting these results. On the other hand, other sources of bias may arise in real data, from unaccounted scale-dependent effects. For instance, the intrinsic alignment of galaxies can introduce correlations between the galaxy shapes that are scale-dependent and may impact cosmic shear measurements \cite{harnois2022cosmic}. Neutrino masses can also introduce supression of the matter power spectrum on small scales and impact the non-Gaussian statistics \cite{Coulton:2018ebd, 2023arXiv231208450B}, influencing the overall cosmological constraints. Given the multitude of potential scale-dependent effects, it is crucial to carefully account for and model all relevant astrophysical and other systematics when conducting cosmological analyses. Our main results may also indicate that other systematic uncertainties might play a more significant role in shaping the cosmological constraints obtained from the real HSC Y1 observational data.

\section{\label{sec: conclusions} Conclusions}

In this paper, we investigate the impact of baryons on the convergence power spectrum and non-Gaussian statistics: peak counts, minimum counts, PDF and scattering transform using HSC Y1-like lensing convergence maps. To quantify the effect of baryons we consider  convergence maps built upon cosmological hydrodynamical simulations $\kappa$TNG; and BAHAMAS low-AGN, fiducial-AGN, and high-AGN. The impact of each of these baryonic feedback scenarios is then propagated into cosmological constraints on $S_8$ and $\Omega_m$ for HSC Y1 mock data. 

Furthermore, in this work we report constraints on cosmological parameter $S_8$ for HSC Y1 real data using all non-Gaussian statistics listed above, acccounting for various analysis choices.   Our main results and conclusions are:

\begin{itemize}
    \item We find that the impact of baryons on the statistics evolves with redshift, as shown Fig. \ref{fig:ratios1} and Fig. \ref{fig:ratios2}. This effect is larger for the third tomographic bin in all the statistics considered, in agreement with previous works \cite{harnois2022cosmic}. In general, our results show that the BAHAMAS high-AGN feedback model produces the strongest impact of baryonic effects on all summary statistics for HSC Y1 mock data, also consistent with other non-Gaussian statistics works \cite{Osato:2020sxo}. 

    \item The impact of baryons on the constraints on $S_8$ for the analysis of mock data is shown in Fig. \ref{fig:ratios3}. The general tendency is that the baryonic feedback models shift $S_8$ towards lower values ($\Delta S_8 < 0$), being more significant for the smallest scales (2 arcmin, $\ell > 900$). The baryonic feedback model BAHAMAS low-AGN tends to shift $S_8$ in the other direction, namely $\Delta S_8 > 0$ for PDF and minimum counts, however, it is statistically non-significant. This is due to the enhancement of structures in the intermediate to low $\kappa$ regime when considering this feedback scenario. 
      
    \item We observe that all non-Gaussian statistics are impacted by baryons at a similar level. Based on the BAHAMAS high-AGN model, the peak counts, PDF and scattering transform exhibit a $0.91\sigma$ to $0.98\sigma$ bias on $S_8$ when considering the smallest scales (2 arcmin, $\ell_{\text{center}}=1200$). The minimum counts reach a $-0.55\sigma$ bias for the same scale. These biases are mainly driven by the differences in constraining power of the statistics employed. For the 5 arcmin scale, the biases are $-0.47\sigma$ for peak counts and $-0.84\sigma$ for the ST, with PDF and minimum counts showing biases $\sim-0.3\sigma$. For the BAHAMAS fiducial-AGN, biases are less than $-0.45\sigma$.


    \item We perform parameter inference using HSC-Y1 real data under the different scenarios we considered for the simulations. This is shown in Fig.~\ref{fig:realdata}. Among the non-Gaussian statistics, the PDF and scattering transform yield the tightest constraints. A similar result was found in \cite{2023A&A...675A.120E}, running a Fisher forecast analysis instead of a full MCMC inference pipeline.  

    \item For real data, we do not find any significant shifts in $S_8$ when incorporating the small scales, with a $\Delta S_8$ much less prominent than the results found for mock data. Hence, our analysis suggests that weak baryonic scenarios are more likely to reproduce HSC-Y1 real data systematics at small scales or that other scale-dependent systematics  may have a more substantial impact in constraining the cosmological parameters. Our results are consistent with other Stage-III analyses using the two-point information~ \cite{huang2021dark, 2023MNRAS.526.5494M, 2022MNRAS.516.5355A}.
    \item As a final remark, all $S_8$ constraints found in this work are consistent with \textit{Planck} TT+TE+EE+lowE result $S_8 = 0.834 \pm 0.016$ \cite{Planck:2018vyg} and the recent Planck data release PR4 analyses $S_8=0.819\pm 0.014$ \cite{2024A&A...682A..37T,2022MNRAS.517.4620R}.
\end{itemize}


This work highlights the advantages of exploring various non-Gaussian statistics and the impact of systematics to perform weak lensing analyses. By considering various statistical measures beyond the power spectrum, we can gain a more comprehensive understanding of how baryonic physics impacts cosmological parameter estimation, in addition to providing improved constraining power on $S_8$.
 
We expect future analyses will be able to discriminate more precisely between baryonic feedback prescriptions and determine if the lower amplitude in $S_8$ found in some LSS studies can be attributed to baryonic feedback systematics. The non-Gaussian statistics are particularly valuable for this task, as they provide crucial insights into the impact of different feedback models on cosmological constraints.

Upcoming Stage-IV surveys will provide precise constraints on cosmological parameters, and thus the biases produced by unmodelled baryonic physics will be more significant than the ones we found in this work. For such surveys, scale cuts based on the analysis of baryon-corrected data vectors with varying baryonic feedback strength need to be revisited.\\

\begin{acknowledgments}
We thank Joaquín Armijo, Will Coulton, Alex Drlica-Wagner, Kevin M. Huffenberger, Xiangchong Li, Ken Osato, and Elena Sellentin for useful discussions.
We thank the organizers of the Kyoto CMBxLSS workshop, as part of the discussions and results that took place during the workshop.
DG and LT would like to thank Kavli IPMU for the hospitality where
part of this work was conducted. GM is part of Fermi Research Alliance, LLC under Contract No. DE-AC02-07CH11359 with the U.S. Department of Energy, Office of Science, Office of High Energy Physics. DG acknowledges support by the Doctorado Nacional/2019-2119188 and L'Oréal-UNESCO fellowship. This work was supported by JSPS KAKENHI Grants 23K13095 and  23H00107 (to JL), 19K14767 and 20H05861 (to MS). We acknowledge support from the MEXT Program for Promoting Researches on the Supercomputer Fugaku hp230202 (to JL). LT acknowledges support by the UTokyo-Princeton Strategic Partnership Teaching and Research Collaboration. SC acknowledges the support of the Martin A. and Helen Chooljian Member Fund and the Fund for Natural Sciences at the Institute for Advanced Study. 
\end{acknowledgments}

\bibliography{GrandonBaryons}

\end{document}